\documentclass[aps,prl,reprint,showpacs,showkeys,groupedaddress,nofootinbib]{revtex4-1}
\usepackage{amsmath,amssymb,amsfonts}
\usepackage{graphicx}
\usepackage{txfonts}
\usepackage{epstopdf}
\usepackage[T1]{fontenc}
\usepackage{subfigure}
\usepackage{float}

\usepackage[unicode]{hyperref}
\hypersetup{
	unicode=true,          
	a4paper=true,
	plainpages=false,
	pdftitle={Title of PDF},    
	pdfauthor={Author of PDF},     
	pdfsubject={Subject of PDF},   
	colorlinks=true,       
	linkcolor=blue,          
	citecolor=blue,        
	filecolor=blue,      
	urlcolor=blue           
}
\newcommand{\refeq}[1]{Eq.~(\ref{#1})}

\newcommand{\reffig}[1]{Fig.~\ref{#1}}

\begin{document}

\author{Kui-Tian Xi$^{1}$}
\author{Hiroki Saito$^{2}$}
\affiliation
{$^1$Beijing National Laboratory for Condensed Matter Physics, Institute of Physics, Chinese Academy of Sciences, Beijing 100190, People's Republic of China\\
$^2$Department of Engineering Science, University of Electro-Communications, Tokyo 182-8585, Japan}

\title{Droplet formation in a Bose-Einstein condensate with strong dipole-dipole interaction}

\begin{abstract}
Motivated by the recent experiment [H. Kadau \textit{et al.}, arXiv:1508.05007 (2015)], we study roton instability and droplet formation in a Bose-Einstein condensate of $^{164}$Dy atoms with strong magnetic dipole-dipole interaction. We numerically solve the cubic-quintic Gross-Pitaevskii equation with dipole-dipole interaction, and show that the three-body interaction plays a significant role in the formation of droplet patterns. We numerically demonstrate the formation of droplet patterns and crystalline structures, decay of droplets, and hysteresis behavior, which are in good agreement with the experiment. Our numerical simulations provide the first prediction on the values of the three-body interaction in a $^{164}$Dy Bose-Einstein condensate. We also predict that the droplets remain stable during the time-of-flight expansion. From our results, further experiments investigating the three-body interaction in dipolar quantum gases are required.
\end{abstract}

\maketitle

Dipolar Bose-Einstein condensates (BECs) of atoms with large magnetic dipole moments, such as chromium \cite{crbec2005}, dysprosium \cite{dybec2011}, and erbium \cite{erbec2012}, are systems in which the long-range and anisotropic dipole-dipole interaction strongly affects their static and dynamic properties. The researches on such a dipolar system both in theories and in experiments are driven by the search for new novel phases in condensed matter physics. Structured ground states and roton excitation spectrum in a pancake-shaped trap have been studied \cite{structured1,structured2,structured3,structured4,structured5,structured6,structured7}. Anisotropic expansion \cite{expansion2005} and collapsing instability \cite{lahaye2007,pressure,lahaye2008} have been observed in a chromium BEC. Increasing attention has also been focusing on bright solitons \cite{soliton1,soliton2}, anisotropic superfluidity \cite{bohn2010,bohn2011}, Faraday patterns \cite{santos2010}, and multicomponent BECs \cite{saito2009,ktx2011,bohn2012,multi4}.
A binary BEC with a strong dipole-dipole interaction
exhibits instability and forms patterns similar to those in magnetic
liquids, such as hexagonal, soliton-like, and labyrinthine
patterns \cite{saito2009}. Droplet formation has also been investigated in dipolar atomic systems \cite{droplet1,droplet2}.

In the recent experiment reported in Ref.\cite{pfau2015}, interaction-induced periodic patterns spontaneously formed in a BEC of dysprosium atoms. By using Feshbach resonance to control the ratio between the s-wave and dipole-dipole interactions, they observed discrete droplet patterns arranged in a long-lived triangular lattice. This result indicates possibility that the system possesses a stable periodic state with matter-wave coherence, which is therefore a candidate of supersolidity \cite{supersolid1,supersolid2,supersolid3}. Before exploring this possiblity, a theoretical understanding of the experimental results in Ref.\cite{pfau2015} is required.

In this Rapid Communication, we propose a theoretical model to explain the experimental results in Ref.\cite{pfau2015}. One finds that the standard mean-field Gross-Pitaevskii model with dipole-dipole interaction cannot reproduce the experimental results; the strong Roton instability is always followed by the d-wave dipolar collapse, which hinders the droplet formation. To circumvent this problem, we propose to include the three-body interaction, which provides an extra replusion to stop the dipolar collapse and stabilizes the droplets. From our numerical simulations, in the presence of three-body interaction, a strong dipolar BEC develops discrete droplet structures, which is in good agreement with the experiment. The three-body interaction in a BEC has been studied theoretically \cite{3body1,3body2,3body3,3body4,3body5} and observed in the recent experiment \cite{3body2015}. Additionally, three-body interaction can stablize the supersolid states in two-dimensional dipolar bosons \cite{3bodydipolar}. We hope our work here in this Rapid Communication will inspire further researches on three-body interaction in a dipolar BEC. 

We consider a dipolar BEC described by the time-dependent macroscopic wave function $\psi(\mathbf{r},t)$ in the zero temperature mean-field approximation. The wave function obeys an extended nonlocal cubic-quintic Gross-Pitaevskii equation (GPE) given by 
\begin{eqnarray}\label{gpe}
\left( i - \gamma \right)\hbar \frac{\partial \psi(\mathbf{r}, t)}{\partial t} &=& \Bigg[ -\frac{\hbar^{2} \nabla^{2}}{2m} + V\left(\mathbf{r}\right) + G \Big| \psi(\mathbf{r}, t) \Big|^{2} + G_{3} \Big| \psi(\mathbf{r}, t) \Big|^{4} \nonumber\\
& & + \int U\left(\mathbf{r} - \mathbf{r'}\right) \Big| \psi(\mathbf{r'}, t) \Big|^{2} d \mathbf{r'}\Bigg] \psi(\mathbf{r}, t),
\end{eqnarray}
where $m$ is the mass of a $^{164}$Dy atom and the trap potential is $V(\mathbf{r}) = m ( \omega_{x}^{2} x^{2} + \omega_{y} y^{2} + \omega_{z}^{2} z^{2} ) / 2$ with $\omega_{x}$, $\omega_{y}$ and $\omega_{z}$ the trap frequencies in $x$, $y$, and $z$ directions, respectively. The wave function is normalized by the number of atoms $N$ as $\int |\psi(\mathbf{r}, t)|^2 d\mathbf{r} = N$. The two-body and three-body interaction coefficients are denoted by $G = 4 \pi \hbar^{2} a / m$ and $G_{3}$, where $a$ is the s-wave scattering length. The atomic spin is fully polarized in the $z$ direction and the dipolar interaction has the form $U ( \mathbf{r} ) = \mu_{0} \mu^{2} ( 1 - 3 z^{2} / r^{2}) / ( 4 \pi r^{3} )$, where $\mu_{0}$ is the magnetic permeability of vacuum and $\mu = 9.93 \mu_{B}$ is the magnetic dipole moment of $^{164}$Dy atom with $\mu_{B}$ the Bohr magneton. The parameter $\gamma$ introduces the energy dissipation to enhance the droplet pattern formation, which has been used to describe, e.g., the dynamics of the vortex lattice formation \cite{dissipation}.

In the experiment \cite{pfau2015}, the s-wave scattering length is tuned to $a \approx a_{dd}$ using a magnetic field in the vicinity of a Feshbach resonance to obtain a stable $^{164}$Dy BEC, where $a_{dd} = \mu_{0} \mu^{2} m / ( 12 \pi \hbar^{2} )$ is a length scale characterizing the magnetic dipole-dipole interaction. For the experiment, the dipolar length $a_{dd} = 132 a_{0}$ is used, where $a_{0}$ is the Bohr radius. The $^{164}$Dy BEC is trapped in a pancake-shaped trap with harmonic frequencies of $(\omega_{x}, \omega_{y}, \omega_{z}) = 2\pi \times (46, 44, 133)$. Subsequently, the magnetic field is tuned to reduce $a$ to the background scattering length $a_{bg} < a_{dd}$, which results in the formation of droplet patterns arranging in ordered structures.

We first prepare the stationary state of the system, obtained by imaginary-time propagation ($i - \gamma = -1$) of \refeq{gpe} with $a = a_{dd}$, which corresponds to the process of getting a stable BEC in the experiment. The numerical simulation is performed using psedospectral method and the dipolar term is calculated using a fast Fourier transform. Subsequently, we use the result of imaginary-time propagation with small initial noise as an initial state of the following real-time propagation. The s-wave scattering length is suddenly changed from $a = a_{dd}$ to $a_{bg}$ at $t = 0$. When $\gamma \neq 0$, we enforce the normalization of the wave function in every time step.

In our calculations, we take $a_{bg} = 71.8 a_{0}$ to reproduce the experimental results \cite{pfau2015}, which is smaller than the measured value $a_{bg} = 92(8) a_0$ \cite{feshlev2015,feshpfau2015}. The scattering length may be affected by the dense spectrum of Feshbach resonances \cite{pfaumfvalue2015}. The three-body interaction coefficient $G_{3}$ is a complex number with $Re[G_{3}]$ describing the three-body scattering parameter and $Im[G_{3}]$ describing the three-body recombination loss \cite{3body2015}. In our simulations, we take $Re[G_{3}] = 3.3 \times 10^{-27} \hbar$ $cm^{6}/s$ and $Im[G_{3}] = -6 \times 10^{-30} \hbar$ $cm^{6}/s$, which are $1 / 100$ of those measured in the experiment of $^{85}$Rb near Feshbach resonance \cite{3body2015}. The value of the three-body interaction coefficient $G_{3}$ depends on the atomic species and is also affected by Feshbach resonance. It is therefore difficult to predict the precise value of $G_{3}$ \cite{g3note}, and we tune $G_{3}$ to fit our numerical results with the experimental results. $Re[G_{3}]$ determines the density peaks of the droplets, which are related to the number of droplets. $Im[G_{3}]$ is chosen to reproduce the atomic decay in the experiment. The scaling property is characterized by dimensionless parameters, $\omega_{y} / \omega_{x}$, $\omega_{z} / \omega_{x}$, $aN / l_{x}$, $G_{3}N^{2} / (\hbar \omega_{x} l_{x}^{6})$, $\mu_{0} \mu^{2} N / (\hbar \omega_{x} l_{x}^{3})$ with $l_{x} = \sqrt{\hbar / (m\omega_{x})}$.

\begin{figure}[t]
	\begin{center}
		\includegraphics[width=0.47\textwidth]{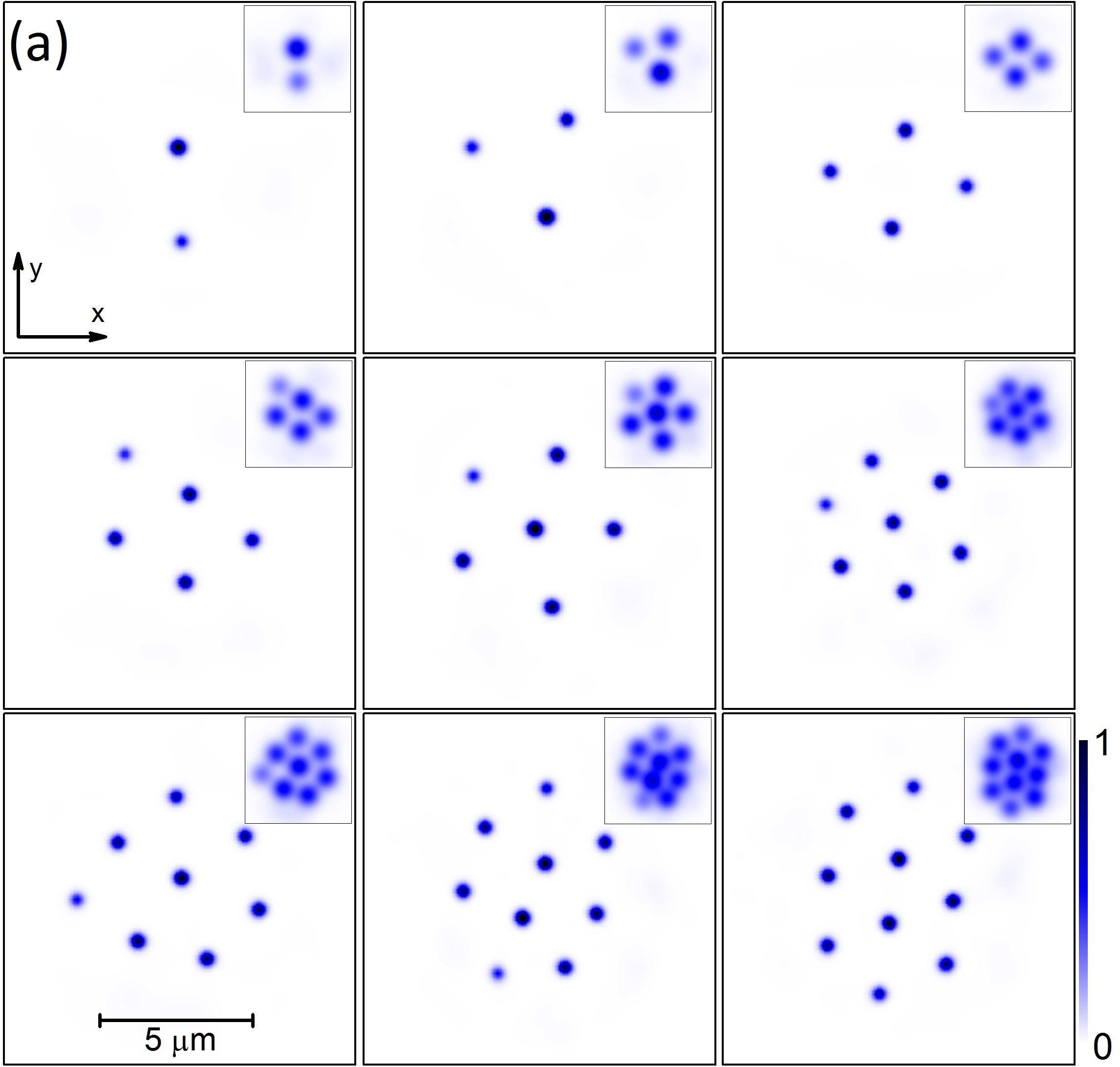}\
		\includegraphics[width=0.47\textwidth]{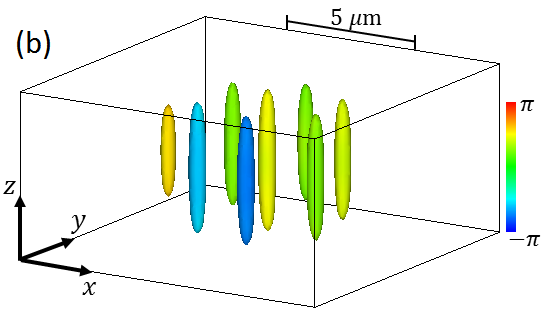}
	\end{center}\vspace{-0.5cm}
	\caption{\label{fig:Fig1} (Color online) Droplet patterns with droplet numbers $N_{d}$ ranging from 2 to 10. (a) Column density profiles at $t = 10$ - $20$ ms after a sudden quench of two-body interaction. In the insets, we take into account one micrometer resolution as in the experiment by taking the convolution of the column density profiles with the Gaussian with one micrometer width. The field of view is $11.5 \times 11.5$ $\mu m$ as well as in the insets. The unit of the density is $N / l_{x}^{2}$. (b) The typical isodensity surface with $N_{d} = 8$. The color represents the phase at the surface. The size of the box is $11.5 \times 11.5 \times 5.8$ $\mu m$.}
\end{figure}

In \reffig{fig:Fig1}, we show droplet patterns formed in the dynamics, where we take $\gamma = 0.006$ which is chosen to reproduce the experimental results. If $\gamma$ is much larger, the droplets merge with each other to reduce the energy of the system, and the number of droplets become smaller than that observed in the experiment; If $\gamma$ is smaller, it takes a longer time to form the triangular patterns. We examined various values of $N$ and obtained the droplets ranging from 2 to 10 ordered in the triangular lattice, as observed in the experiment \cite{pfau2015}. By smoothing the column density profile using the experimental resolution, the images become similar to those in the experiment (insets in \reffig{fig:Fig1}(a)). These ordered triangular lattice of droplets are obtained as a metastable state. A typical isodensity surface with $N_d = 8$ is plotted in \reffig{fig:Fig1}(b), showing that the droplets exhibit cigar shapes. Each droplet has the phase coherence along the long axis, whereas different phases are distributed to the droplets.

\begin{figure}[t]
	\begin{center}
		\includegraphics[width=0.47\textwidth]{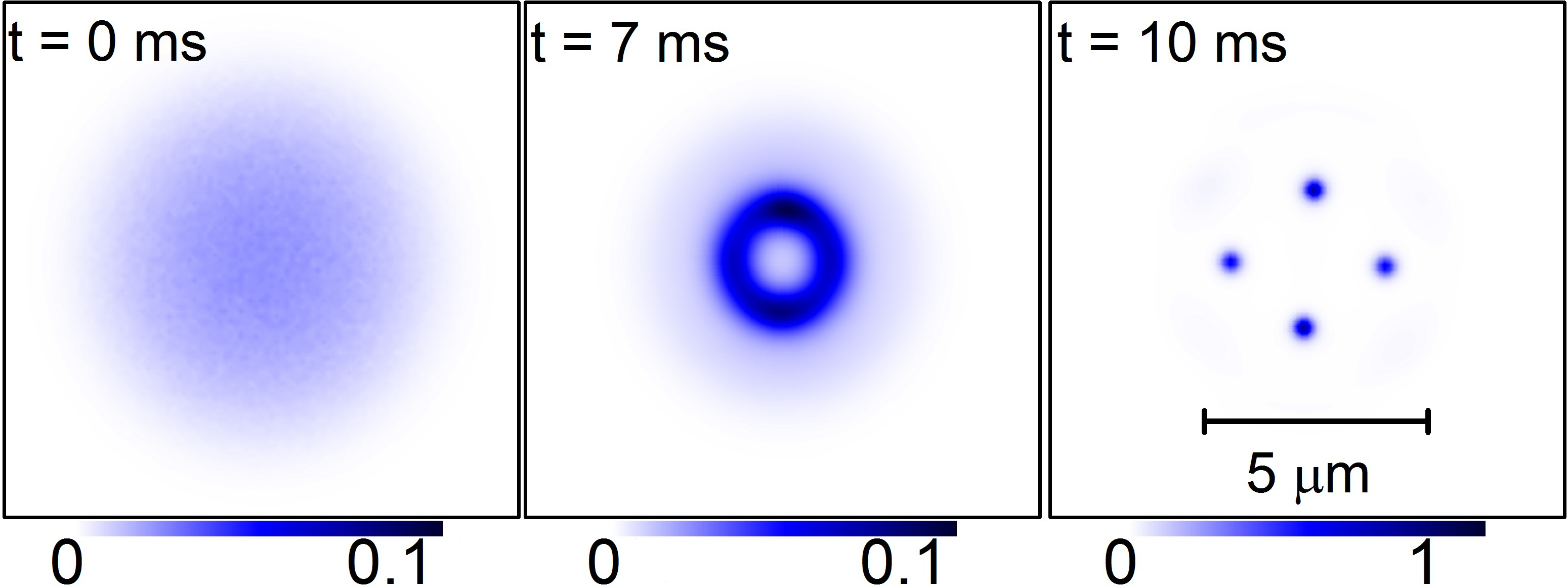}
	\end{center}\vspace{-0.5cm}
	\caption{\label{fig:Fig2} (Color online) Column density profiles in the dynamics of 4-droplet formation. The number of atoms is $N = 7500$. The unit of the density is $N / l_x^2$ and the field of view is $11.5 \times 11.5$ $\mu m$.}
\end{figure}

We also plot the typical dynamics of the formation of 4 droplets in \reffig{fig:Fig2}. A ring pattern develops at $t = 7$ ms, which is followed by the breaking into droplets. The dynamics of the system after the quench of the two-body interaction is as follows. Initially the quantum pressure and the repulsive s-wave interaction can overcome the attractive part of the dipolar interaction to stabilize the BEC in a harmonic trap \cite{lahaye2007,pressure}. Decreasing the s-wave scattering length, the dipolar attraction becomes dominant, and the BEC starts to collapse, during which the three-body repulsion stops the collapse and the condensate splits into droplets. The three-body repulsion can stop the collapse and stabilize the droplets, since the dipolar and three-body energies are proportional to $(d_{\perp}^{2} d_{z})^{-1}$ and $(d_{\perp}^{2} d_{z})^{-2}$, respectively, where $d_{\perp}$ and $d_{z}$ are the radial and axial sizes of the droplet. If the three-body repulsion was absent, density peaks would undergo violent collapse and explosion \cite{lahaye2008}, whereas no atomic burst was observed in the experiment \cite{pfau2015}. Immediately after the droplets are formed, they sometimes exhibit regular patterns reflecting the symmetry of the system, which however does not settle into the crystal structure without the energy dissipation $\gamma$. The cigar-shaped droplets aligned side by side repel each other because of anisotropy of the dipolar interaction, and they form the triangular lattice when the excess energy is dissipated.

\begin{figure}[t]
	\begin{center}
		\includegraphics[width=0.47\textwidth]{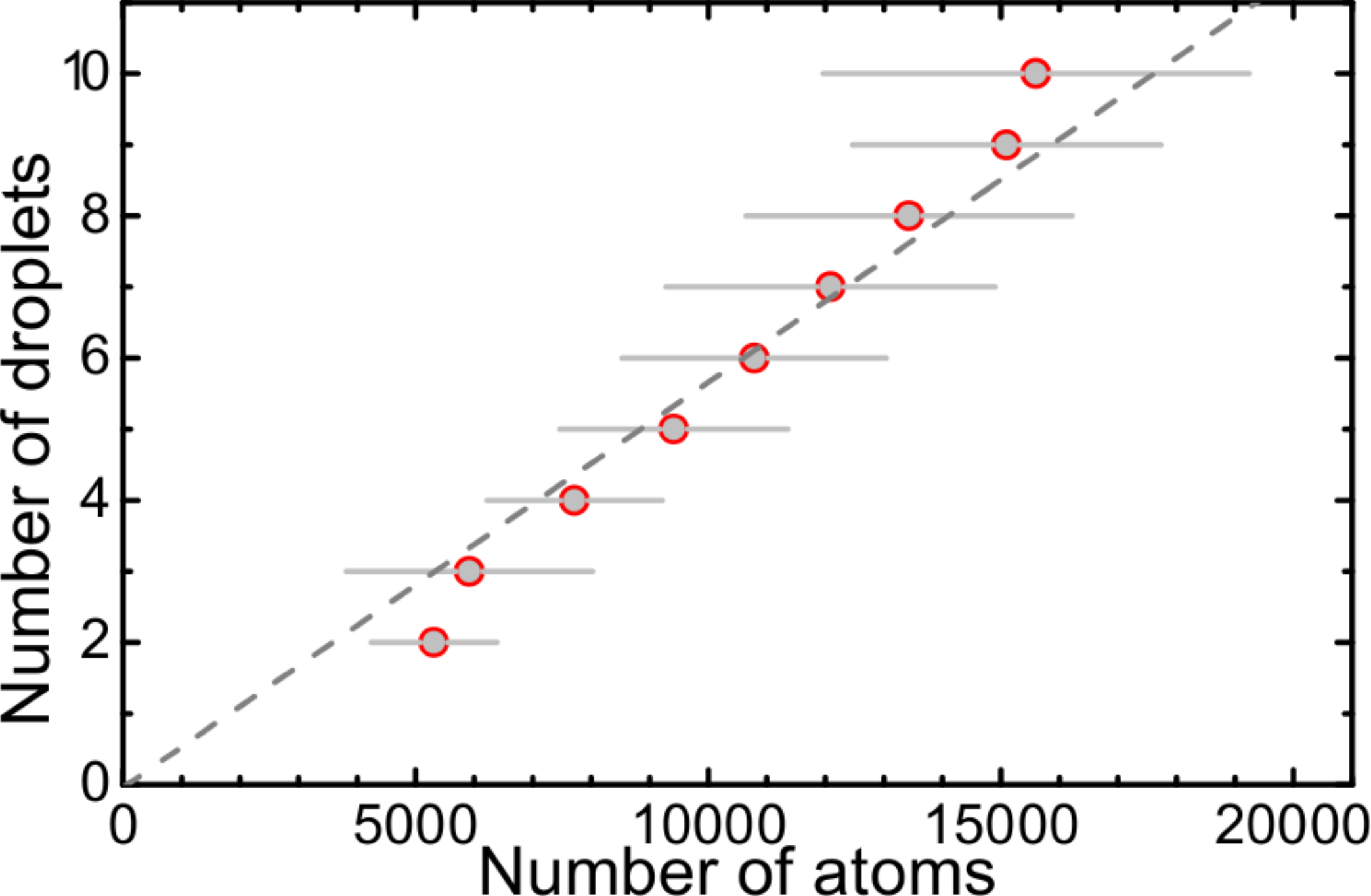}
	\end{center}\vspace{-0.5cm}
	\caption{\label{fig:Fig3} (Color online) Relation between number of droplets $N_{d}$ and number of atoms $N$. We perform 280 calculations with different numbers of atoms and different initial noises for statistical analysis. The error bars show the standard deviation. The dashed line indicates $N_{d} = N / 1755$.}
\end{figure}

We then perform 280 runs of numerical simulations with different numbers of atoms and different initial noises for statistical analysis, and plot the relation between $N_{d}$ and $N$ in \reffig{fig:Fig3}, which approximately have a linear dependence as in the experiment. From the fitted line in \reffig{fig:Fig3}, we identify a slope of about 1755 atoms per droplet, which is in good agreement with the experimental result indicating a slope of 1750(300) atoms per droplet \cite{pfau2015}.

\begin{figure}[t]
	\begin{center}
		\includegraphics[width=0.47\textwidth]{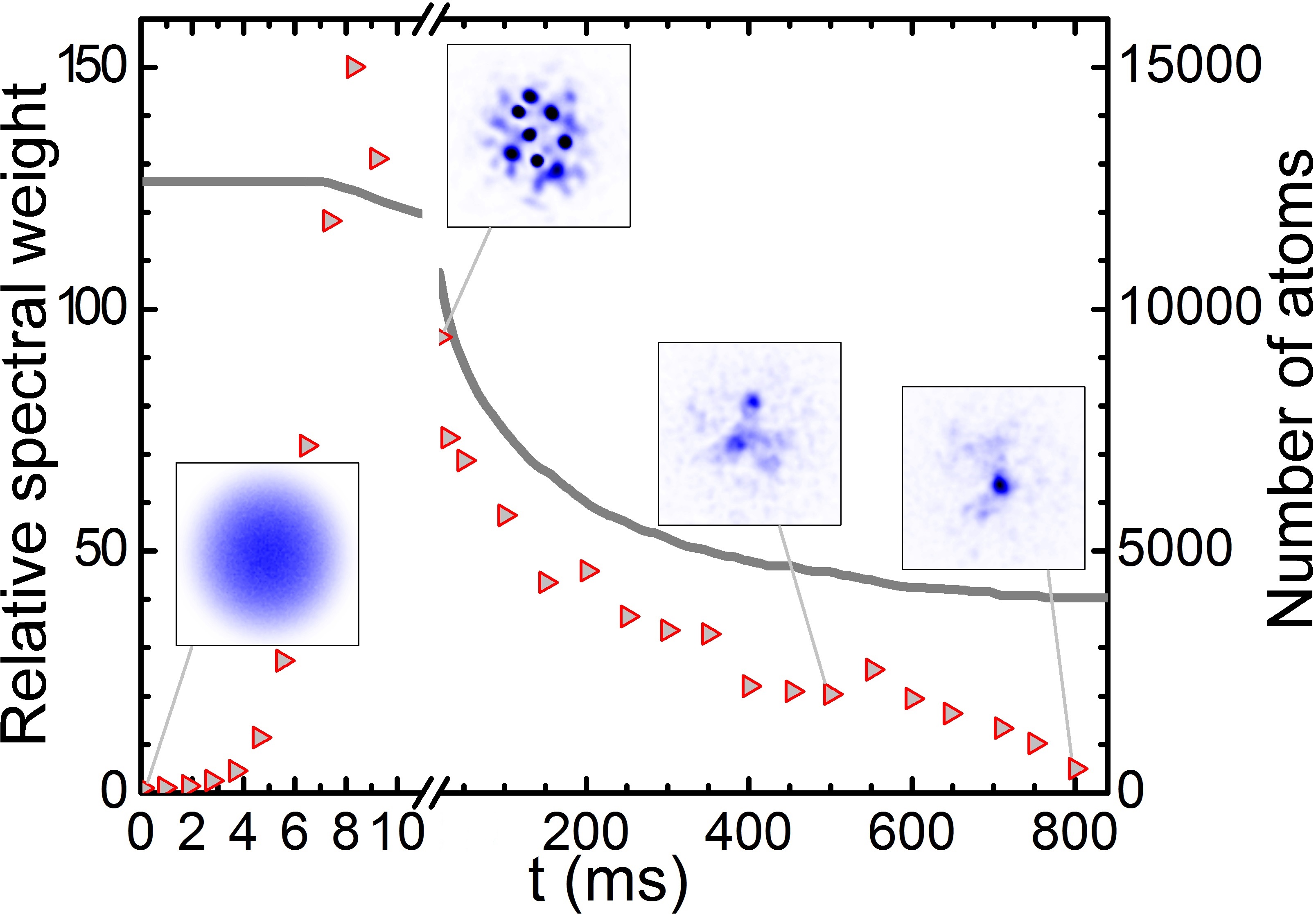}
	\end{center}\vspace{-0.5cm}
	\caption{\label{fig:Fig4} (Color online) Time evolution of the relative spectral weight (red triangles) and the number of atoms (grey solid line). The insets show the column density profiles at $t = 0$, 20, 500, and 800 ms. The field of view in the insets is $11.5 \times 11.5$ $\mu m$.}
\end{figure}

We subsequently investigate the formation time and lifetime of droplet patterns (see \reffig{fig:Fig4}). The droplet structures decay with atom loss, and it is shown that the droplets merge together until reaching a single droplet with a decrease in the number of atoms. The lifetime of the droplet patterns is up to 500 ms, which is in good agreement with the experiment \cite{pfau2015}. We find that the main reason for the reduction of atomic lifetime is the three-body recombination loss enhanced by the formation of the droplets in which the atomic density is large. In fact, the atomic loss prominently occurs after the formation of the droplets \cite{pfau2015}. We take $\gamma = 0$ in this and the following simulations.

Two dimensional Fourier spectrum $S(k_{x},k_{y})$ of the column density profile is computed for further analysis. We quantify the degree of ordering as \cite{pfau2015}
\begin{equation}
SW(t) = \sum_{k = 1.5 \, \mu m^{-1}}^{5 \, \mu m^{-1}} S(k_{x}, k_{y}, t)
\end{equation}
with $k^{2} = k_{x}^{2} + k_{y}^{2}$. The relative spectral weight is defined as $SW(t) / SW(0)$, which is plotted in \reffig{fig:Fig3}. The value of the relative spectral weight in \reffig{fig:Fig3} is much larger than that in the experiment, since $SW(0)$ in our ideal initial state is very small, while $SW(0)$ in the experiment is increased by environmental noises.

\begin{figure}[t]
	\begin{center}
		\includegraphics[width=0.47\textwidth]{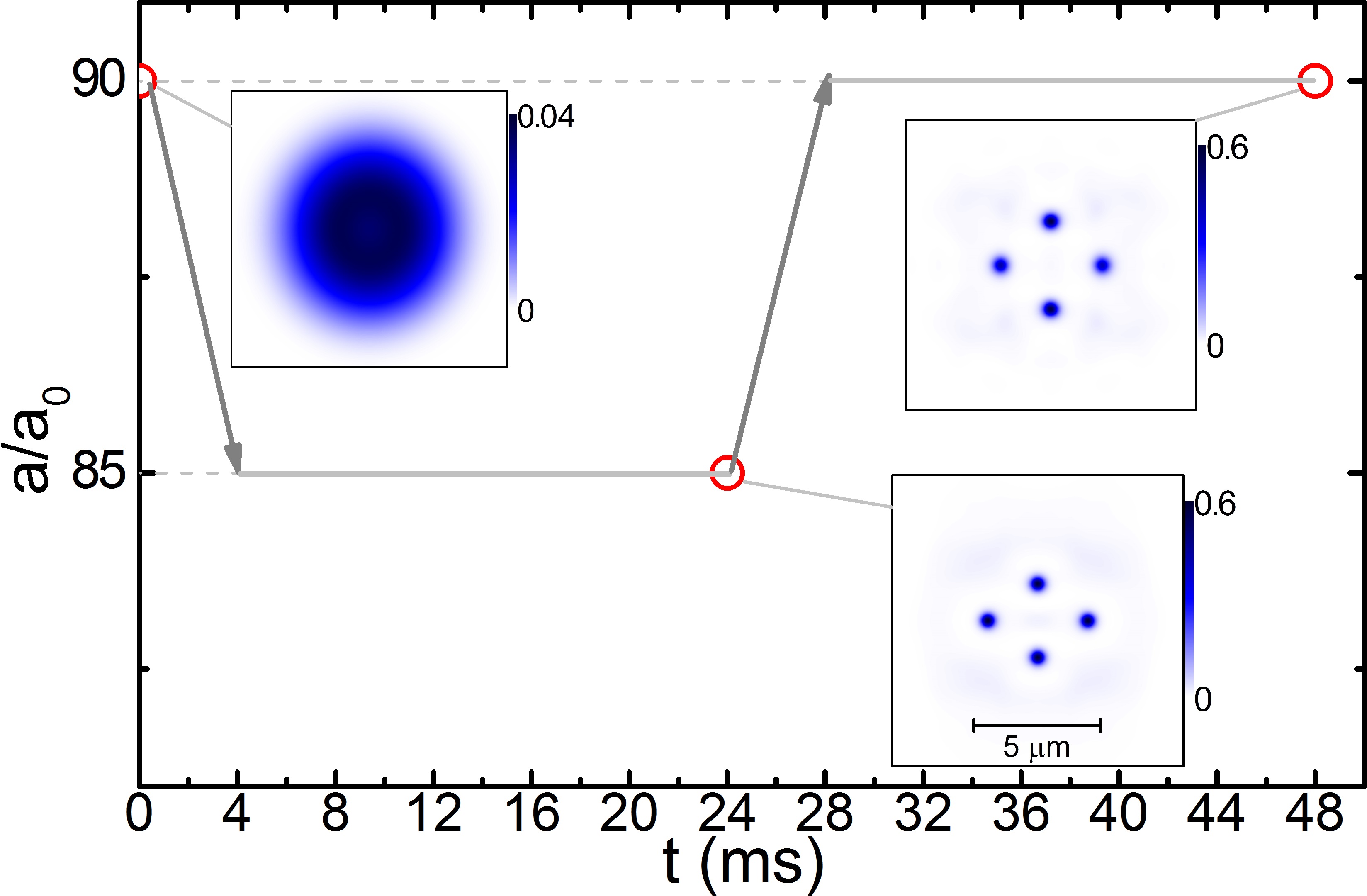}
	\end{center}\vspace{-0.5cm}
	\caption{\label{fig:Fig5} (Color online) Hysteresis of pattern formation. We prepare a stable state at $a = 90a_{0}$, then perform time evolution with reducing $a$ to $85a_{0}$ in 4 ms. Subsequently, we keep $a = 85a_{0}$ for 20 ms, then increase $a$ back to $90a_{0}$ in 4 ms and keep it for 20 ms. The insets show the column density profiles at $t = 0, 24$, and 40 ms, where the unit of the density is $N / l_{x}^{2}$ and the field of view is $11.5 \times 11.5$ $\mu m$.}
\end{figure}

We then perform the following numerical sequence to identify the hysteresis and bistability, which is illustrated in \reffig{fig:Fig5}. The initial state is the stationary state with $N = 15000$ for $a = 90 a_{0}$, and $a$ is reduced to $85 a_{0}$ in 4 ms. After holding $a$ unchanged for 20 ms, we increase $a$ back to $90 a_{0}$. The result in \reffig{fig:Fig5} indicates that the Thomas-Fermi profile is stable for $90 a_{0}$ as $a$ is decreased from above, whereas the droplet pattern remains for $90 a_{0}$ as $a$ is increased from $85 a_{0}$, showing the hysteresis behavior as in the experiment \cite{pfau2015}.

The true ground state in the presence of the three-body repulsion is a single droplet that contains all the atoms. In our simulation, however, the multiple-droplet structures are obtained as metastable states dynamically. In the experiment, on the other hand, the multiple droplets are also obtained by evaporative cooling with $a \approx a_{bg}$, implying that the multiple-droplet state is the ground state. This seeming contradiction can be resolved by the following scenario. When the condensate grows in the evaporative cooling and exceeds the critical number for the collapse, the dynamic droplet formations occur, which result in multiple droplets. Once such a multiple-droplet state is produced, it is metastable and survives for a long time.

\begin{figure}[t]
	\begin{center}
		\includegraphics[width=0.47\textwidth]{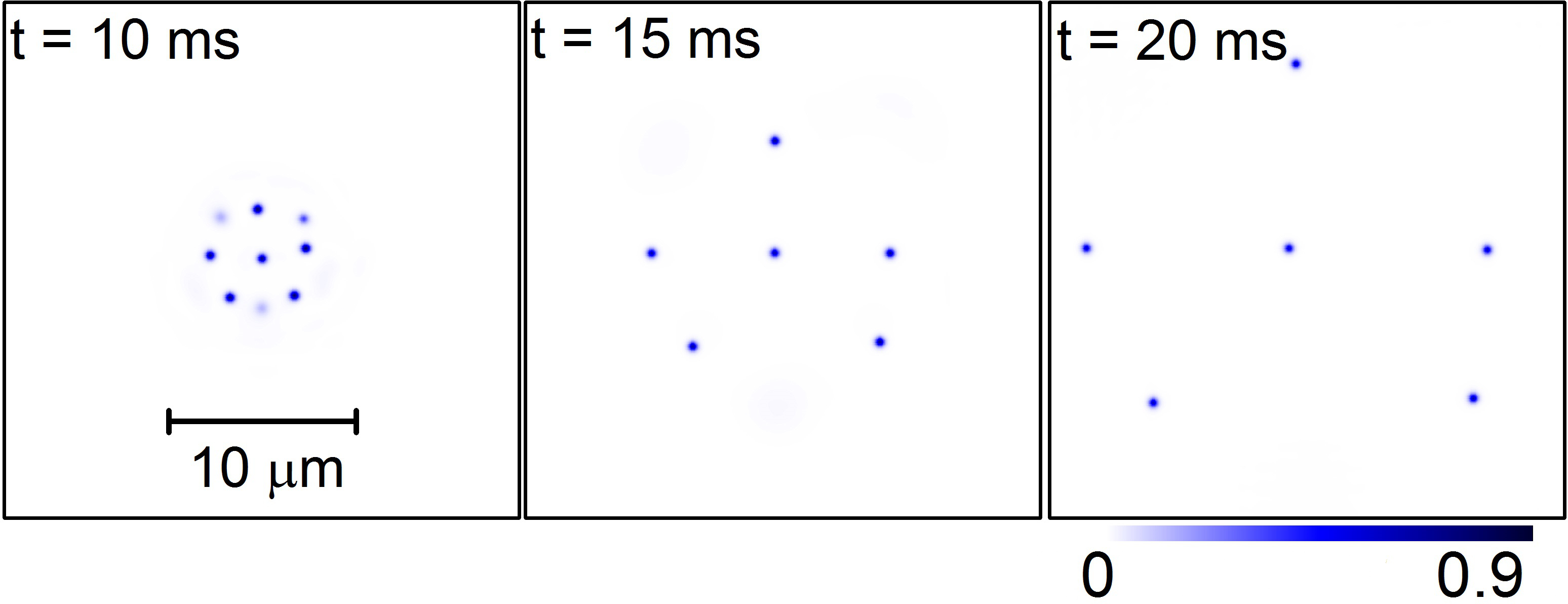}
	\end{center}\vspace{-0.5cm}
	\caption{\label{fig:Fig6} (Color online) Column density profiles in a time-of-flight simulation, where the trap potential is switched off at $t = 10$ ms. The number of atoms is $N = 15000$. The unit of the density is $N / l_x^2$ and the field of view is $25 \times 25$ $\mu m$.}
\end{figure}

Finally, we perform a time-of-flight (TOF) simulation, where the trap potential is switched off at $t = 10$ ms. The column density profiles at $t = 10$ms, $15$ms, and $20$ms are shown in \reffig{fig:Fig6}. The result indicates that the droplet pattern expands after the switch-off of the trap potential, while the size of each droplet remains unchanged. This is a theoretical prediction which can be easily confirmed with the current experimental setup in Ref.\cite{pfau2015} by switching off the optical dipole trap, followed by the absorption imaging. If the droplet stabilization and pattern formation are caused by another mechanism, such as quantum many-body effect, the droplets will disappear during the TOF. This is because an individual droplet, or a bright soliton, should be unstable against collapse or expansion in a three-dimensional free space without the three-body repulsion.

In conclusion, we have investigated the droplet pattern formation in a $^{164}$Dy BEC with dipolar and three-body interactions. We found that in the presence of the three-body interaction, the results of our numerical simulations are in good agreement with the recent experiment reported in Ref.\cite{pfau2015}. As in the experiment, droplet patterns form with different number of droplets ranging from 2 to 10, which linearly increases with the number of atoms. We showed that the droplet structures decay due to the three-body recombination loss. The hysteresis behavior of the system was also shown. To date, there are no other theoretically or experimentally precise values of coefficients of the three-body interaction in a $^{164}$Dy BEC, so the good agreement of our numerical results with the experiment indicates that our calculations give the first prediction on the values of the three-body interaction in a BEC of dysprosium atoms. Moreover, we predict that the droplet pattern will expand during the TOF while each droplet remains its size, which can be identified in future experiments. It is clear from our results that further experiments studying three-body interaction in dipolar quantum gases are required.

\textit{Acknowledgements.}
K.-T. Xi would like to thank S. Yi for helpful discussion and X. Cui for kind support. This work was supported by NSFC 11374177, JSPS KAKENHI Grant Number 26400414, and MEXT KAKENHI Grant Number 25103007.

\textit{Note added.} Very recently, a preprint \cite{Blakie2015} appeared, which also employs the three-body interaction to account for the droplet formation in a dysprosium BEC.


\begin{thebibliography}{99}
\bibitem{g3note}
According to Refs.\cite{3body2,3body3,3body4}, the value of $G_3$ may be proportional to $16\pi(4\pi-3\sqrt{3}) \hbar^2 a^4/m$, which is $\sim 10^{-29} \hbar$ $cm^6/s$ in the present parameters.
\end{thebibliography}

\end{document}